\documentclass[%
 reprint,
 amsmath,amssymb,
 aps,
 prresearch,
 floatfix,
]{revtex4-2}
\usepackage{empheq} 
\usepackage{xcolor}
\usepackage{graphicx}
\usepackage{dcolumn}
\usepackage{bm}
\usepackage{hyperref}
\usepackage{comment}
\usepackage{cases}
\usepackage{braket}
\hypersetup{
    colorlinks=true,
    linkcolor=blue,
    citecolor=blue,
    filecolor=blue,
    urlcolor=blue
}
\makeatletter
\let\revtex@onecol\onecolumngrid
\let\revtex@twocol\twocolumngrid
\makeatother
\begin{document}
\preprint{APS/123-QED}

%\title{Energy resolution of Rydberg states at the two-photon thermal limit}
\title{Fundamental linewidth limit of electromagnetically induced transparency in a thermal Rydberg ladder}

\author{Noah~Schlossberger}
 \email{noah.schlossberger@nist.gov}
\affiliation{National Institute of Standards and Technology, Boulder, Colorado 80305, USA}
\author{ Nikunjkumar~Prajapati}
\affiliation{National Institute of Standards and Technology, Boulder, Colorado 80305, USA}
\author{Alexandra~B.~\mbox{Artusio-Glimpse}}
\affiliation{National Institute of Standards and Technology, Boulder, Colorado 80305, USA}
\author{Samuel~Berweger}
\affiliation{National Institute of Standards and Technology, Boulder, Colorado 80305, USA}
\author{Christopher~L.~Holloway}
\affiliation{National Institute of Standards and Technology, Boulder, Colorado 80305, USA}

\date{\today}% It is always \today, today,

\begin{abstract}
Spectroscopy of Rydberg states has become a popular platform for quantum sensing, with the most common readout scheme being two-photon electromagnetically induced transparency (EIT) using counter-propagating laser beams. In this scheme, the energy resolution of the Rydberg state is set by the spectral linewidth of the EIT feature. While selection criteria for the two-photon resonance can narrow the linewidth to the order of the Rydberg state decay rate for a single atom, the Doppler shift from thermal velocity of the atoms broadens the ensemble linewidth to the order of the decay rate of the intermediate state. Here, we derive an analytic expression for the Doppler residual lineshape in the low-power limit and corroborate the results with experiment. For Rb, we find the full-width at half-maximum linewidth limit to be 1.84 MHz when scanning the coupling laser and measure an experimental linewidth of 2.04~MHz. These linewidths are around a factor of two narrower than previous theoretical estimates as well as previously reported measured linewidths. With this, we demonstrate the most precise two-photon energy resolution of a Rydberg state in thermal vapor to date. We then map out broadening mechanisms near this limit.
\end{abstract}

\maketitle

\begin{figure}[h]
\centering
\includegraphics[scale = .9]{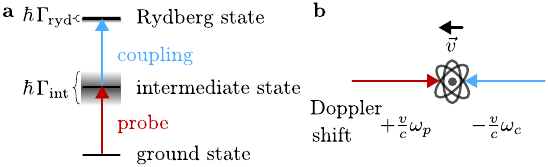}
\caption{Two-photon EIT configuration. a) The energy level diagram. b) The beam orientations and the resulting Doppler shift on each beam for a moving atom.}
\label{fig:ELD_beamconfig}
\end{figure}

\paragraph*{Introduction.}
Rydberg states are a powerful platform for quantum sensing of electromagnetic radiation and fields \cite{Adams_2020,9374680,Schlossberger2024}, 
with applications in electric field imaging \cite{10.1063/1.4883635,Fan:14,PhysRevA.92.063425,PhysRevX.10.011027,Schlossberger:25,sp9j-ps66}, 
communications 
\cite{9069423,8778739,jlrg-6889}, 
radar 
\cite{10.1063/5.0287757, chen2025highresolutionquantumsensingrydberg} 
and source localization 
\cite{6dl6-754w}, 
thermometry 
\cite{PhysRevLett.107.093003,
PhysRevApplied.23.044037, 
PhysRevResearch.7.L012020},  
%beierle2026developmentquantumblackbodythermometer
%eckel2026atomic
and SI-traceable metrology of fields and voltage \cite{6910267,10.1063/1.5045212,
9363580,
10.1116/5.0090892}. 
By far the most common method to probe the energy of Rydberg states is with two-photon electromagnetically induced transparency (EIT) in a ladder scheme \cite{Sedlacek2012}. In this scheme, the two lasers are counter-propagating to cancel out the Doppler shift that arises from thermal motion of the atoms. However, the frequency mismatch of the lasers leads to imperfect cancellation. Because of this, the width of the two-photon spectroscopic lineshape is dominated by the ``Doppler residual.'' While EIT has been studied for decades \cite{PhysRevLett.66.2593} and a number of resources exist on the subject \cite{RevModPhys.77.633, Finkelstein_2023}, this Doppler residual lineshape for Rydberg states in a two-photon ladder EIT configuration has yet to be properly theoretically characterized. Initial estimates of the linewidth provided without derivation in Refs.~\cite{Daschner2015Addressable,10.1117/12.2309386} have influenced physical assumptions in a large body of work \cite{ PhysRevApplied.20.L061004, 10.1063/5.0147827, ZHANG20241515, Schmidt:24, PhysRevApplied.23.034028, 10722127, 10.1117/12.2586718, 10.1117/12.2616797, duverger2024metrologie}. In this manuscript, we demonstrate that these estimates overestimate the linewidth by a factor of around two. Here, we derive a simple analytic form for the Doppler residual lineshape (Eq.~\ref{eq:dopplerlineshape}) and demonstrate this lineshape experimentally. This represents the fundamental limit to the linewidth of two-photon EIT spectroscopy in a thermal vapor, and thus the limit on the energy resolution of the Rydberg state.
\paragraph*{Theoretical linewidth.}
The energy level diagram and beam configuration for the two-photon EIT readout of Rydberg states are demonstrated in Fig.~\ref{fig:ELD_beamconfig}.
The so called ``Doppler residual'' linewidth arises from the decay width of the intermediate state, mediated by a small subset of Doppler classes of atoms. The linewidth predicted by Refs.~\cite{Daschner2015Addressable,10.1117/12.2309386} can be derived from conservation of energy. For the two photon process to occur, both photons must lie within the energy uncertainty of each state in the atom's frame (Fig.~\ref{fig:ELD_beamconfig}a). Because the intermediate and Rydberg states decay at a rate $\Gamma$, their energy has a full width at half maximum (FWHM) uncertainty of $\hbar \,\Gamma$ where $\hbar$ is the reduced Planck constant (via Heisenberg uncertainty or Fourier analysis).  In the lab frame, the detuning of each laser is modified by the respective Doppler shift (Fig. \ref{fig:ELD_beamconfig}b). This means
\small
\begin{equation}
\begin{cases}
\left|(\Delta_p + \frac{v}{c}\omega_p) + (\Delta_c  - \frac{v}{c} \omega_c)\right| \leq \frac{\Gamma_{\text{ryd}}}{2} & \begin{cases}\text{probe+coupling= }\\
\text{Rydberg energy}\end{cases}\\
\left|\Delta_p + \frac{v}{c} \omega_p\right| \leq  \frac{\Gamma_{\text{int}}}{2} & \begin{cases}\text{probe = }\\
 \text{intermediate energy,} \end{cases}
\label{eq:restframe}
\end{cases}
\end{equation}
\normalsize
where $\omega_p$ and $\omega_c$ are the angular frequencies of the probe and coupling lasers, $\Delta_p$ and $\Delta_c$ are the angular frequency detunings of the probe and coupling laser, $v$ is the velocity of the atom in the laser beams' axis, and $c$ is the speed of light. Because the Rydberg state decay rate is negligible compared to that of the intermediate state, we can set $\Gamma_{\text{ryd}} \rightarrow 0$ and solve the first inequality in Eq.~\ref{eq:restframe} (as an equality) to find the velocity. Plugging this into the second inequality in Eq.~\ref{eq:restframe} and simplifying, we get
\begin{equation}
\left| \Delta_p\frac{\omega_c}{\omega_c - \omega_p} + \Delta_c\frac{\omega_p}{\omega_c - \omega_p}\right| \leq \frac{\Gamma_{\text{int}}}{2} .
\label{eq:rescondition}
\end{equation}
Interpreting these equations as bounds on the detunings, we can obtain the angular FWHM linewidth $\Gamma_\textrm{EIT}$ of the two-photon resonance from this equation:
\begin{equation}
    \Gamma_\text{EIT} = \Gamma_\text{int} \cdot \begin{cases}\frac{\omega_c-\omega_p}{\omega_c}  & \text{scanning probe}\\
    \frac{\omega_c-\omega_p}{\omega_p}  & \text{scanning coupling}
    \end{cases}.\label{eq:original_FWHM}
\end{equation}

This result matches the claims of Refs.~\cite{Daschner2015Addressable,10.1117/12.2309386}, predicting a Doppler-residual-limited linewidth of $2\pi \times 3.79$~MHz for Rb and $2\pi \times 3.51$~MHz for Cs when scanning the coupling laser.
However, this equation does not take into account the non-linearity of the two photon coherence. We find that the actual FWHM is narrower by a numerical factor nearly equal to a factor of two, meaning the actual linewidth limit is given by $2\pi \times1.84$~MHz for Rb and $2\pi \times 1.71$~MHz for Cs. 

A proper treatment of the system requires solving the master equation. The evolution of the density matrix $\rho$ under a Hamiltonian $H$ is given by the Maxwell-Bloch equation:
\begin{equation}
    \frac{d\rho}{dt} = - \frac{i}{\hbar}[H, \rho] + \mathcal{L}[\rho], 
\label{eq:densitymatrix}
\end{equation}
where $\mathcal{L}$ is the Liouvillian superoperator that describes the decoherence of the system. The Hamiltonian for the two level ladder is
\begin{equation}
H
=
\hbar
\begin{pmatrix}
0 & \dfrac{\Omega_p}{2} & 0 \\
\dfrac{\Omega_p^*}{2} & -\Delta_p' & \dfrac{\Omega_c}{2} \\
0 & \dfrac{\Omega_c^*}{2} & -(\Delta_p'+\Delta_c')
\end{pmatrix},\label{eq:Hamiltonian}
\end{equation}\\
where $\Omega_p$ and $\Omega_c$ are the Rabi rates of the probe and coupling lasers and $\Delta'_p$ and $\Delta'_c$ are the probe and coupling laser detunings in the atom's frame.
 These are related to the lab frame detunings $\Delta_p$ and $\Delta_c$ by the Doppler shift:
\begin{equation}
    \begin{cases}
        \Delta'_p = \Delta_p + \frac{v}{c}\omega_p\\
        \Delta'_c = \Delta_c  - \frac{v}{c}\omega_c.
    \end{cases}\,\,\,\, .
\end{equation}
The Liouvillian superoperator is given by
\small
\begin{equation}
\mathcal{L}[\rho]
=
\Gamma_{\mathrm{int}}
|1\rangle\langle2|\rho|2\rangle\langle1|
-\frac{\Gamma_{\mathrm{int}}}{2}\left\{|2\rangle\langle2|,\rho\right\}
-\frac{\Gamma_{\mathrm{ryd}}}{2}\left\{|3\rangle\langle3|,\rho\right\},
\end{equation}
\normalsize
where $\ket{1}$, $\ket{2}$, and $\ket{3}$ are the ground, intermediate, and Rydberg states.
This is set up and solved in Ref. \cite{PhysRevA.51.576}. 
Under the weak probe approximation,
\small
\begin{equation}
    \rho_{21}=\frac{ - i \Omega_p}{\frac{\Gamma_\text{int}}{2} -  i(\Delta_p+\omega_p \frac{v}{c}) +\left( \frac{\Omega_c^2/4}{\frac{\Gamma_\text{ryd}}{2}- i(\Delta_p + \Delta_c + (\omega_p - \omega_c)\frac{v}{c})}\right)}.
\end{equation}
\normalsize
The susceptibility of the atoms to the probe light is given by
\begin{equation}
    \chi = \frac{4 \hbar \mu_{21}^2}{\epsilon_0 \Omega_p} \rho_{21}\,\,,
\end{equation}
where $\mu_{21}$ is the transition dipole matrix element and $\epsilon_0$ is the permittivity of free space. The observed susceptibility is then an integral over the velocity distribution $N(v)$ of the atoms:
\begin{equation}
    \chi = \int_{-\infty}^\infty N(v)\chi(v) dv,\label{eq:doppint}
\end{equation}
with the atomic velocity distribution given by 
\begin{equation}
N(v) = \frac{N_0}{v_\text{th} \sqrt{\pi}} e^{-(v/v_\text{th})^2},\text{ with }v_\textrm{th} \equiv  \sqrt{\frac{2k_B T}{m}}.
\end{equation}
%\begin{equation}
%\begin{cases}
%N(v) = \frac{N_0}{v_\text{th} \pi} e^{-(v/v_\text{th})^2}\\
%\text{where}\\
%v_\textrm{th} \equiv  \sqrt{\frac{2k_B T}{m}}
%\end{cases}\,\,\,\, ,
%\end{equation}
Here, $N_0$ is the total atomic number density, $k_B$ is the Boltzmann constant, $T$ is the temperature of the atoms, and $m$ is the mass of one atom.
The result of the integral in Eq.~\ref{eq:doppint} is% (Ref. \cite{PhysRevA.51.576}, Eqs. 10-12)
\fontsize{8.65pt}{9pt}\selectfont
\begin{widetext}
\begin{empheq}[left=\empheqlbrace]{alignat=2}
    \chi \;\;\;=\;&  i\frac{2  \hbar c \mu_{21}^2 N_0 \sqrt{\pi}}{\epsilon_0 \omega_p v_\text{th}}\Bigg(-(1-d) e^{-z_1^2}(1+ \text{erf}(iz_1)) + (1+d) e^{-z_2^2}(1- \text{erf}(iz_2))\Bigg)\label{eq:chi}
\\
    \text{where}\notag\\
    d \;\;\;\equiv\;& \frac{i}{z_1 - z_2}\left(\frac{\frac{\Gamma_\text{int}}{2}-i\Delta_p}{\omega_p v_\text{th}/c}-\frac{\frac{\Gamma_\text{ryd}}{2} - i(\Delta_p + \Delta_c)}{(\omega_p - \omega_c)v_\text{th}/c}\right)\label{eq:d}\\[10pt]
    z_{1,2} \equiv\;& -\frac{i}{2}\left(\frac{\frac{\Gamma_\text{int}}{2}-i\Delta_p}{\omega_p v_\text{th}/c}+\frac{\frac{\Gamma_\text{ryd}}{2} - i(\Delta_p + \Delta_c)}{(\omega_p - \omega_c)v_\text{th}/c}\right) %\notag\\& \quad\quad\quad\quad
    \pm \frac{i}{2} \Bigg(\left(\frac{\frac{\Gamma_\text{int}}{2}-i\Delta_p}{\omega_p v_\text{th}/c}-\frac{\frac{\Gamma_\text{ryd}}{2} - i(\Delta_p + \Delta_c)}{(\omega_p - \omega_c)v_\text{th}/c}\right)^2
    - \frac{{\Omega_c}^2}{\omega_p (\omega_p - \omega_c)v_\textrm{th}^2/c^2}\Bigg)^{1/2}_.
    \label{eq:z}
\end{empheq}
\end{widetext}
\normalsize
\begin{figure*}
\includegraphics[scale = .9, trim = {0 .00cm 0 .01cm}, clip]{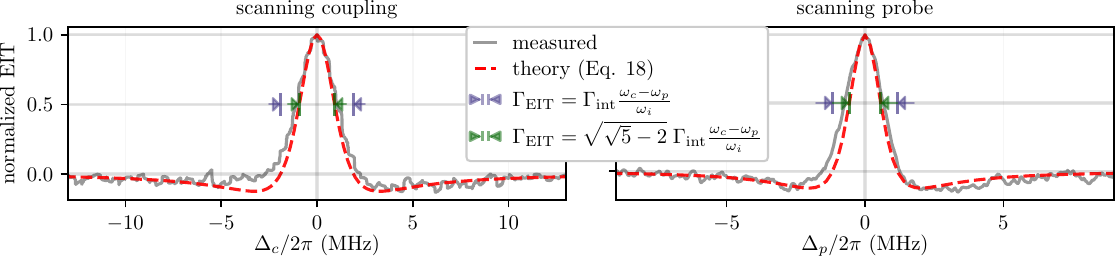}
\caption{Experimental lineshapes of the two-photon EIT resonance when scanning the coupling (left) or probe (right) lasers using the $5S_{1/2}(F=3) \rightarrow 5P_{3/2}(F=4) \rightarrow 48S_{1/2}$ ladder in $^{85}$Rb. The theoretical lineshape given by Eq.~\ref{eq:analtyicImChi} is overlayed in red, with only the amplitude floated between the theory and data. We also display the FWHM predicted by Eq.~\ref{eq:original_FWHM} ($2\pi\times3.79$~MHz scanning $\Delta_c$ and $2\pi\times2.33$~MHz scanning $\Delta_p$) and by Eq.~\ref{eq:EITLW} ($2\pi\times1.84$~MHz scanning $\Delta_c$ and $2\pi\times1.13$~MHz scanning $\Delta_p$).}
\label{fig:theorylineshapes}
\end{figure*}
This equation, while quite general, is hard to interpret. We can simplify with some approximations:
\begin{itemize}
\item 
$\left|\frac{\frac{\Gamma_\text{ryd}}{2}}{(\omega_p - \omega_c) v_\text{th}/c}\right| \ll \left| \frac{\frac{\Gamma_\text{int}}{2}}{\omega_p v_\text{th}/c}\right|$.\\[5pt]
The decay rate of the Rydberg state is very small compared to the intermediate state, so we can take $\frac{\Gamma_\text{ryd}}{2} \rightarrow 0$ in Eqs. \ref{eq:d} and \ref{eq:z}.
\item $|z_{i}|<<1$.\\[5pt]
Because $\{\frac{\Gamma_\text{int}}{2}, \Delta_p, \Delta_c ,\Omega_c\} <<  v_\textrm{th}/c \cdot \{\omega_p, \omega_p - \omega_c \}$, all numerators in Eq.~\ref{eq:z} are small compared to the denominators, so $|z_i| \ll 1$ and we can take  $\text{erf}(z_i)\rightarrow 0$ and the $e^{-z_i^2}\rightarrow 1$ in Eq.~\ref{eq:chi}. Note that here, $\omega_p v_\text{th}/c$ represents the average single-photon angular Doppler shift, on the order of $2\pi\times$100s of MHz. This approximation is essentially that the EIT only talks to a narrow portion of the velocity distribution, and the distribution is essentially flat over this region. Taking this approximation gets rid of the broad Doppler background profile one would observe when scanning the probe.

\end{itemize}
We can then simplify $\chi$ to
\begin{equation}
\chi \propto 
i \sqrt{\frac{1}{\frac{\Omega_c^2 \omega_p  (\omega_c -\omega_p )}{(\frac{\Gamma_\text{int}}{2} (\omega_c -\omega_p )-i (\Delta_p  \omega_c +\Delta_c  \omega_p ))^2}+1}} .
\end{equation}
The absorption of the probe light is proportional to the imaginary part of the susceptiblity. We can find the imaginary part using a few complex identities:
\begin{multline}
    \text{Im}(\chi) \propto  
    \\
    \frac{1}{\sqrt{\frac{\Omega_c ^2 \omega_p  (\omega_c -\omega_p ) \left((\frac{\Gamma_\text{int}}{2 })^2 (\omega_c -\omega_p )^2-(\Delta_p  \omega_c +\Delta_c  \omega_p )^2\right)}{\left((\frac{\Gamma_\text{int}}{2})^2 (\omega_c -\omega_p )^2+(\Delta_p  \omega_c +\Delta_c  \omega_p )^2\right)^2}+1}}.
    \label{eq:analtyicImChi}
\end{multline}
We can expand this in $\Omega_c/\Gamma_\text{int}$:
\small
\begin{multline}
    \text{Im}(\chi) \propto
   1 \\
   -\Omega_c^2\frac{ \omega_p (\omega_c-\omega_p) \left(\left(\frac{\Gamma_\text{int} }{2}\right)^2 (\omega_c-\omega_p)^2-(\Delta_p \omega_c+\Delta_c \omega_p)^2\right)}{\left(\left(\frac{\Gamma_\text{int} }{2 }\right)^2 (\omega_c-\omega_p)^2+(\Delta_p \omega_c+\Delta_c \omega_p)^2\right)^2}
   \\ + \mathcal{O}(\frac{\Omega_c}{\Gamma_\text{int}})^4.\label{eq:lowcouplingimchi}
\end{multline}
\normalsize
In the low coupling power limit, we can neglect terms that scale as $(\frac{\Omega_c}{\Gamma_\text{int}})^4$.  Then the EIT signal is given by the second term in Eq.~\ref{eq:lowcouplingimchi}. The strength of the EIT signal is proportional to $\Omega_c^2$, which is proportional to the power of the coupling laser. The lineshape of the EIT feature can be written as
\begin{equation}
\begin{cases}
   \text{lineshape} = \frac{1-\eta^2}{(1+\eta^2)^2} \\[5pt]
    \eta \equiv \frac{\Delta_p \omega_c + \Delta_c \omega_p}{(\Gamma_\text{int}/2)(\omega_c - \omega_p)}
    \end{cases}.
    \label{eq:dopplerlineshape}
\end{equation}
Note that this lineshape is purely Lorentzian, it contains wings arising from the $ \frac{1-\eta^2}{(1+\eta^2)^2}$ form that go negative and converge towards zero quadratically. The FWHM linewidth of Eq.~\ref{eq:dopplerlineshape} in angular frequency is
\begin{equation}
    \Gamma_\text{EIT} = \sqrt{\sqrt{5}-2} \;\;\Gamma_\text{int} \cdot \begin{cases}\frac{\omega_c-\omega_p}{\omega_c}  & \text{scanning probe}\\
    \frac{\omega_c-\omega_p}{\omega_p}  & \text{scanning coupling}
    \end{cases}\,\,\, .
    \label{eq:EITLW}
\end{equation}
This is a factor of $\sqrt{\sqrt{5}-2} \approx 0.486$ lower than Eq.~\ref{eq:original_FWHM}, giving lineshapes more than a factor of two narrower than the previous estimate of the fundamental limit.
\paragraph*{Experimental lineshapes.}
We compare the lineshapes given by Eq.~\ref{eq:dopplerlineshape} to experimentally measured lineshapes for $^{85}$Rb ($\Gamma_{int} = 38.1(1)$~MHz \cite{PhysRevA.66.024502}, $\omega_p = \frac{2\pi c}{780\text{ nm}}$, $\omega_c = \frac{2\pi c}{480\text{ nm}} $) in Fig.~\ref{fig:theorylineshapes}. Our experimental FWHM lineshapes when scanning the coupling and probe are $2\pi \times$2.04(9)~MHz and $2\pi \times$1.38(4)~MHz, respectively. Note that the energy resolution of the Rydberg state corresponds to the linewidth when scanning the coupling laser. While the feature is narrower by a factor of $f_p/f_c$ when scanning the probe, any shift to the Rydberg energy also gets scaled by this same factor when observed through the EIT feature, mediated by the relative Doppler shifts between the two lasers \cite{Schlossberger2024}. Our Rydberg energy resolution  FWHM is thus $\hbar \times 2\pi \times 2.04(9)$~MHz, which is 10\% above the fundamental limit. 

\begin{figure}
    \includegraphics[scale = .9, trim = {0 .2cm 0 .2cm}, clip]{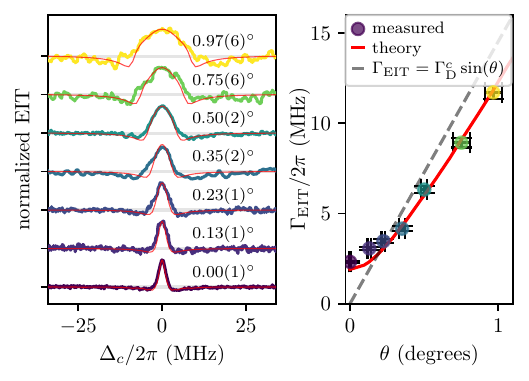} 
    \caption{Misalignment broadening of the two-photon EIT lineshape. Left: Measured lineshapes at various angles, with the numeric evaluation of Eq.~\ref{eq:misalignmentchi} overlayed in red. Right: The extracted linewidths are plotted as a funciton of the misalignment angle.}
    \label{fig:beam_misalignment_measurement}
\end{figure}

\paragraph*{Misalignment broadening.}
The main barrier to achieving the lineshape given by Eq.~(\ref{eq:dopplerlineshape}) is mismatch of the alignment of the $k$-vectors of the laser beams, which can significantly increase the EIT linewidth \cite{Su:25} by coupling an orthogonal velocity component into one of the beams' Doppler shifts.
Let $\theta$ be the angle between the coupler and the probe beams. If we define our coordinate system such that $v_\parallel$ is the velocity of the atom along the probe beam and $v_\perp$ is the velocity of the atom orthogonal to the probe (in the plane defined by the two beams), then we can write the Doppler shifts as
\begin{equation}
\begin{cases}
    \Delta'_p = \Delta_p +\frac{v_\parallel}{c}f_p \\
    \Delta'_c = \Delta_c -\frac{v_\parallel\cos(\theta)}{c}f_c -\frac{v_\perp\sin(\theta)}{c}f_c
    \end{cases}\,\,\,\, .
\end{equation}
We can then plug these into Eq.~\ref{eq:Hamiltonian} and, solve Eq.~\ref{eq:densitymatrix}, and integrate over both velocity distributions:
\begin{equation}
    \chi = \int_{-\infty}^\infty \int_{-\infty}^\infty N(v_\parallel)N(v_\perp)\chi(v_\parallel, v_\perp) \;dv_\parallel dv_\perp. \label{eq:misalignmentchi}
\end{equation}
While this is difficult to do analytically, we evaluate this numerically. We compare the expected lineshapes and linewidths via this treatment and experimentally measured lineshapes and linewidths in Fig.~\ref{fig:beam_misalignment_measurement}.
The dashed line indicates a na\"ive expression for the broadening given by the Doppler width from the orthogonal projection, $\Gamma_\text{D}^c \sin(\theta)$, where $\Gamma_\text{D}^c$ is the single-photon absorption Doppler width for the coupling laser \cite{hooker2010laser}: 
\begin{equation}
    \Gamma_\text{D}^c = 2\sqrt{2\ln(2)}\frac{\omega_c}{c}\sqrt{\frac{ k_BT }{m}}.
\end{equation}
From this, we can see that in order to reach the Doppler residual linewidth limit, the $k$-vectors of the probe and coupling beams must be aligned to within $\theta<0.1$~degrees. It is of note that this broadening mechanism preserves the negative wings of the lineshape, whereas most other broadening mechanisms do not. The existence of the wings then serves as a signature that indicates that beam misalignment is the  dominant source of broadening.

\begin{figure}
    \includegraphics[scale=.9, trim = {0, .2cm, 0, .2cm}, clip]{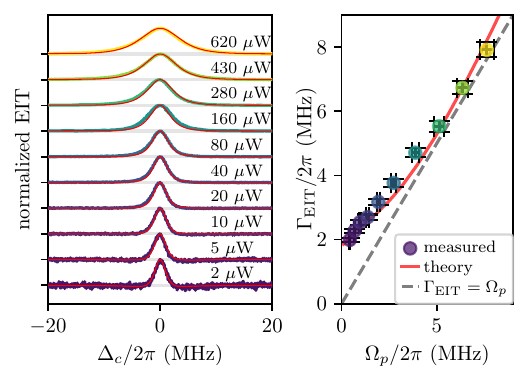}
    \caption{Probe power broadening of the two-photon EIT lineshape. Left: Measured lineshapes at various probe powers, with the numeric evaluation of Eq.~\ref{eq:doppint} overlayed in red. Right: The extracted linewidths are plotted as a function of probe Rabi frequency.}
    \label{fig:probepower}
\end{figure}

\paragraph*{Power broadening.}
Another large barrier to observing the Doppler residual limited linewidth is power broadening from the probe laser. Avoiding probe power broadening requires intensities below the $\mu$W/mm$^2$ level, which for many setups places the signal beneath the noise floor (to observe these lineshapes in Fig.~\ref{fig:theorylineshapes} required lock-in amplification and averaging on the order of 1000 traces). To characterize the effect of power broadening, we compare lineshapes calculated numerically from Eq.~\ref{eq:densitymatrix} to measured lineshapes for a variety of probe Rabi rates in Fig.~\ref{fig:probepower}.
Note that the negative wings disappear after only a MHz or so of power broadening. It is also of interest that the derivative of the power-broadened linewidth does not go to zero at low power, indicating that the scaling does not exhibit traditional linewidth quadrature-sum behavior.
\paragraph*{Other broadening mechanisms.}
In order to observe the Doppler residual linewidth limited lineshape, no other broadening mechanisms may broaden by an amount comparable to the lineshape. In Table~\ref{tab:broadening}, we provide rough calculations of a variety of broadening mechanisms, the resulting bounds on experimental parameters, and the values that we used to take the lineshapes in Fig.~\ref{fig:theorylineshapes}.
\begin{table}[h]
\resizebox{\linewidth}{!}{
\begin{tabular}{l| l|l |l}
mechanism & broadening & requirement ($\ll\sim$1~MHz) & value (Fig.~\ref{fig:theorylineshapes})\\ \hline
beam misalignment & $\sim \Gamma^c_\text{D} \sin\theta$ & $\theta \ll\sim0.07^\circ$ & $<0.03^\circ$\\
power broadening & $\sim \Omega_p$ & $I_p \ll \sim 1.5$~$\mu$W/mm$^2$ & 0.05 $\mu$W/mm$^2$\\
 & $\sim \Omega_c$ & $I_c \ll\sim 200$ mW/mm$^2$ & 1.1 mW/mm$^2$\\
transit time & $\sim v_\textrm{th}/\sigma_\text{beam}$ & $\sigma_\text{beam} \gg\sim  0.02$ mm& 2.5 mm\\
Zeeman splitting & $\sim \mu_\textrm{eff}|B|/\hbar$ & $B \ll\sim30$  $\mu$T & $<5$~$\mu$T\\
DC Stark & $\sim \alpha E^2/\hbar$ & $E \ll\sim0.2$ V/m & unknown
\end{tabular}
}
\caption{Requirements for reaching the Doppler residual linewidth limit in $48S$ state of Rb. $\Omega_p$ and $\Omega_c$ are probe and coupling Rabi rates, $I_p$ and $I_c$ are probe and coupling laser intensities, $\sigma_\text{beam}$ is the FWHM beam diameter, $v_\text{th}$ is the expectation value of the magnitude of the atomic velocity, $\alpha$ is the polarizability of the Rydberg state, and $E$ and $B$ are the background electric and  magnetic field strengths. Dipole moments and polarizabilities are calculated using the ARC python package \cite{SIBALIC2017319}.}
\label{tab:broadening}
\end{table}

Transit time broadening arises from decoherence due to thermal motion of the atoms across the beam. The broadening can be approximated as one over the expectation value of the time an atom spends in the beam. In addition, a background magnetic field lifts the degeneracy of the angular momentum projection states, leading to a broadening. The effect of Zeeman shifts on two photon EIT lineshapes is discussed in detail in Ref. \cite{PhysRevA.109.L021702}. We wrap the vapor cell in high permeability  magnetic shielding foil and find that the magnetic field at the location of the vapor cell is below 5~$\mu$T using a magnetometer. Finally, an inhomogeneous DC electric field can also cause broadening. While we do not have a way to measure the electric field inside of the vapor cell, we believe electric field to be minimum inside the vapor cell due to the shielding from external fields by alkali atoms on the surface of the cell \cite{PhysRevApplied.13.054034}. However, it is known that visible light from the coupling laser can induce charge distributions in the vapor cell due to the photoelectric effect in alkali atoms on the surface of the cell \cite{10.1116/5.0264378}. This may be responsible for the small broadening we observe.

\paragraph*{Methods.}
The probe and coupling lasers were Gaussian beams with FWHM diameters of 2600$\times$2460~$\mu$m and 3380$\times$2420~$\mu$m respectively. Both lasers were locked to an ultra-low expansion cavity using sidebands of an electro-optic modulator (EOM) to allow for an arbitrary offset frequency from the cavity modes. To scan the lasers, we scanned the frequency of the EOMs. This allowed the repeatability required for averaging many scans. The scan range was verified with a precision wavemeter. To achieve adequate SNR, the coupling laser was chopped using an acousto-optic modulator (AOM), and the photodetector output was demodulated using a lock-in amplifier phase-locked to the AOM amplitude. A chop frequency of 33~kHz was chosen to be above the $1/f$ noise of the system but below the bandwidth limit of the atomic response and the photodetector.

To calculate the probe Rabi rate in Fig.~\ref{fig:probepower}, we inferred the peak intensity using the measured total power and the Gaussian widths extracted from a beam profiling camera. Because the probe laser resolved the hyperfine $F$ levels (driving $F=3$ to $F'=4$) but not the projection $m_F$, we took a root-mean-square average of the allowed $m_F$ transitions to find the effective Rabi rate:
\begin{align}
\Omega_p
&=
\frac{E_0}{\hbar}\,
\frac{1}{\sqrt{7}}\,
\left|
\langle 5P_{3/2} \| d \| 5S_{1/2} \rangle
\right|\\
&=2\pi \cdot \sqrt{I} \left(793600 \frac{\text{Hz}}{\sqrt{\text{W/m$^2$}}}\right),\label{eq:rabi}
\end{align}
where the reduced dipole matrix element was taken from Ref. \cite{PhysRevA.92.052501} to be  5.978(4)~$e\cdot a_0$. Because each part of the gaussian beam contributed to the EIT proportional to its intensity, we took the intensity in Eq.~\ref{eq:rabi} to be the mean intensity-weighted intensity, exactly 1/2 the peak intensity for gaussian beams. 

The angle $\theta$ between the beams was measured by measuring the divergence of the beams over a distance of $\approx$2~m, and taking the inverse tangent of the displacement over the measurement distance. For Fig.~\ref{fig:theorylineshapes}, the beams were constrained to within 1~mm over 3~m, bounding the misalignment to $\theta<0.02^\circ$. The beams were also collimated to a divergence of less than $0.03^\circ$.

\paragraph*{Acknowledgements.} \hspace{.23\linewidth}\\The authors thank \mbox{Daniel}~\mbox{Hammerland} and \mbox{Link} \mbox{Patrick} for constructive comments during the internal review of this manuscript.
%\\[5pt]
This research  was supported by NIST under the NIST-on-a-Chip program.  A contribution of the U.S. government, this work is not subject to copyright in the U.S.
%\\[5pt]
%\\[5pt]
The authors declare no conflict of interest.
\paragraph*{Data availability.}
All of the data presented in this paper and used to support the conclusions of this article is available at \cite{MIDAS}.
%\bibliography{main}

\end{document}